\documentclass[english]{article}
\usepackage{mathptmx}
\usepackage{helvet}
\usepackage{courier}
\usepackage[T1]{fontenc}
\usepackage[latin9]{inputenc}
\usepackage[a4paper]{geometry}
\usepackage{amsmath}
\usepackage{graphicx}
\usepackage{setspace}
\makeatletter

\providecommand{\tabularnewline}{\\}

\usepackage{multicol}
\usepackage{fancyhdr}

\makeatother

\usepackage{babel}

\begin{document}

\title{Thermally induced changes of structure in Ni$_{50}$Mn$_{25+x}$Ga$_{25-x}$
magnetic shape memory single crystals with very low twinning stress}

\author{L. Straka$^{1}$%
\thanks{Corresponding author. Email: ladislav.straka@aalto.fi Phone: +358
50 415 2886 %
}, J. Drahokoupil$^{2}$, O. Pacherová$^{2}$, K. Richterová\emph{$^{2}$},
V. Kopecký$^{2}$, \\
H. Hänninen$^{1}$, and O. Heczko$^{2}$}

\maketitle
$^{1}$ Aalto University School of Engineering, Laboratory of Engineering
Materials, PO Box 14200, FIN-00076 Aalto, Finland 

$^{2}$ Institute of Physics ASCR, Na Slovance 1999/2, 182 21 Prague,
Czech Republic \\

\begin{abstract}
In search for the origins\emph{ }of the extraordinary low twinning
stress of Ni-Mn-Ga magnetic shape memory alloys we studied the thermally
induced changes of structure in Ni$_{50}$Mn$_{25+x}$Ga$_{25-x}$
($x$=2.7--3.9) single crystal samples and compared them with twinning
stress dependences. The alloys exhibited transformation to five-layered
(10M) martensite structure between 297 to 328 K. All samples exhibited
magnetic shape memory effect. Just below the transformation temperature
the samples had very low twinning stress of about 0.1--0.3 MPa, which
increased with decreasing temperature. The structural changes were
monitored using X-ray diffraction in the temperature range 173--343\,K\emph{.
}The 10M structure was approximated by monoclinic lattice with the
unit cell derived from the cubic unit cell of the parent L2$_{1}$
phase. With decreasing temperature, the lattice parameters $a$ and
$\gamma$ increased, $c$ decreased, while $b$ was nearly constant.
For $x\leq3.5$, sudden sharp changes in $a$ and $b$ parameters
additionally occurred, resulting in $a=b$ in some regions of the
phase diagram, which might be related to the refinement of twin structure
of 10M martensite on nanoscale. The temperature dependences of lattice
parameter $\gamma$ (and $c$ or $c/a$) correlate well with the temperature
dependences of twinning stress in agreement with the prediction by
a microstructural model of twin boundary motion. On the contrary,
there is no correlation between $(a-b)$ and twinning stress. This
indicates no significant role of $a/b$ twins or laminate in twin
boundary motion mechanism and low twinning stress.
\end{abstract}


\twocolumn

\section{Introduction}

Twinning stress is one of the most important parameter of magnetic
shape memory alloys (MSMAs). Only with very low twinning stress the
MSMAs can exhibit the giant straining in magnetic field mediated by
the motion of martensite twin boundaries, which phenomenon is known
as \emph{magnetic shape memory effect} or \emph{magnetically induced
reorientation} (MIR) of martensite \cite{Webster,Ullakko1996,Buschov_book,MSMPhenomena,Wilson_Review,likhachev2006,Sozinov_NM,Sozinov_14M,karaca2006,Straka_2011Jphys}.
The MIR can be utilized in applications requiring fast actuation with
large strain \cite{Wilson_Review}, while the inverse MIR (modification
of magnetic field by the ferromagnetic twin microstructure rearrangement)
can be used for sensing-type applications or vibrational energy harvesting.
It turns out that for good application performance the twinning stress
must typically be as low as possible, of the order of 0.1 MPa \cite{Straka_2011Jphys,Schmidt},
or around 1 MPa in certain cases \cite{Benedict_2012}. That is up
to three orders lower than the twinning stress of ordinary shape memory
materials \cite{sehitoglu}.

The Ni-Mn-Ga based MSMAs with five-layered (10M) martensite structure
demonstrate very low twinning stress, especially for the composition
Ni$_{50}$Mn$_{25+x}$Ga$_{25-x}$, where $x=$2.7--3.9 \cite{Straka_2011Jphys,Straka_IMT}.
The very low twinning stress of the order of 0.1~MPa or even 0.01~MPa
\cite{Kellis001MPa} is observed with Type 2 martensite twin boundaries
\cite{Jaswon_Type_2,Bilby_Type_2,Mogylny,Nishida_2008,Sozinov_Type2}
in a broad temperature interval including room temperature \cite{Straka_IMT,StrakaT1xT2}.
The Type 2 twin boundaries can form in 10M martensite because of the
non-negligible monoclinicity of the nearly tetragonal lattice. They
connect two martensite variants with different orientation of the
$c$-axis by 180$^{\circ}$ lattice rotation around the twin shear
axis. In contrast, the Type 1 twin boundaries \cite{Jaswon_Type_2,Bilby_Type_2},
connecting the two variants by a simple mirroring of the lattice at
the twinning plane, show in average $\approx$1\,MPa twinning stress
at room temperature. The twinning stress further increases with decreasing
temperature with the rate of about 0.04 MPa/K \cite{StrakaT1xT2,Heczko1.7K}.

The origin of the extraordinary low twinning stress in 10M martensite
and sharply different twinning stress of Type 1 and Type 2 twin boundaries
and twinning stress temperature dependences have not yet been fully
explained, despite of the major significance of the subject for the
whole field of MSMAs.\textbf{ }Utilizing first-princi-ples atomistic
simulations and twin nucleation model based on the Peierls\textendash{}Nabarro
formulation, Wang and Sehitoglu \cite{sehitoglu} predicted twinning
stress of 10M martensite to be 3.5~MPa, which is comparable to experimental
value of $\approx$1~MPa for Type 1 twins. To explain the much lower
twinning stress of Type 2 twins, Faran and Shilo \cite{faran2011}
suggested that a thicker (more diffuse) Type 2 twin boundaries experience
a smaller Peierls energy variation and thus require less driving force
to move. Similar argument was presented by Kaufman et al. \cite{Kaufman_Modulated_Mart}.
Heczko et al. \cite{Heczko_Acta2013}, following reasoning by Salje
and Lee et al. \cite{Salje1,Salje2}, tentatively explained the very
low twinning stress of Type 2 twins by flat potential energy landscape
on an atomic scale.\textbf{ }

Theoretical analysis of Rajasekhara and Ferreira \cite{rajasekhara},
and more detailed analysis of Wang and Sehitoglu \cite{sehitoglu}
and Faran and Shilo \cite{faran2013} show that the twinning stress
depends on the shear modulus, the interplanar spacing between the
twinning planes, and the Burgers vector of the twin dislocations.
The latter two depend on the lattice parameters, and the lattice parameters,
in turn, change significantly with temperature \cite{Lanska_2004,Pagounis_abc,Glavatska_icomat,Glavatsky_magnetic}.
In relation to lattice parameters it is also interesting to note that
Sozinov et al. recently demonstrated that the twinning stress of tetragonal
non-modulated (NM) martensite decreased\textbf{ }significantly when
reducing the $c/a$ ratio, resulting in MIR in NM phase \cite{Sozinov_NM}. 

Seiner et al. \cite{Seiner_Microstructural} suggested that in addition
to atomistic models (as e.g. Ref.~\cite{sehitoglu}), also meso-
and micro-structure should be considered as an important factor influencing
the twinning stress. The particular internal twin microstructure can
both decrease or increase the twinning stress considerably and can
play important role in the different behavior of Type 1 and Type 2
twins.\textbf{ }The developed microstructural model based on elastic
continuum theory shows that especially the monoclinic distortion of
the lattice represented by a difference in lattice parameters $(a-b)$
and the monoclinic angle $\gamma$ can control the twinning stress. 

Thus, from various theoretical analyses and different experiments
it seems that the increase of twinning stress with decreasing temperature
can be related to the changes in lattice parameters. This motivated
the present experimental investigation. It is important to note here
that although the twin boundary kinetics in 10M martensite can depend
strongly on thermal activation, the thermal activation may play no
role in twinning stress \cite{ShiloMatSciTech14}. For example very
low $\approx$0.1--0.3~MPa twinning stress of Type 2 twins down to
1.7 K was reported in Refs.~\cite{Straka_IMT,Heczko1.7K}. If there
is no role of thermal activation, the direct linking of twinning stress
changes with changing lattice geometry or structure becomes highly
relevant.

In this article, we investigate the links between the temperature-related
increase in twinning stress and the lattice parameters using the direct
measurements of both properties on the single crystals exhibiting
MIR. We follow the changes of the structure with decreasing temperature
in the same single crystals which exhibit the twinning stress of $\approx$0.1\,MPa
for Type 2 twins at room temperature. In order to take account of
the effects of twin microstructure on twinning stress property, we
pay a special attention to the changes in lattice monoclinicity, i.e.
to the slight difference between $a$ and $b$ lattice axes and to
the slight deviation of the related angle $\gamma$ from 90$^{\circ}$.
The measured temperature dependences of the lattice parameters and
changes in lattice monoclinicity are compared with the temperature
dependences of twinning stress for Type 1 and Type 2 twin boundaries.
Additionally we found previously unreported changes in structure manifested
as sudden, nonmonotonous changes in $a$ and $b$ lattice parameters.

\section{\label{sec:Experimental}Material and methods}

\begin{table*}[!t]
\caption{Nominal and XRFS-determined composition and transformation temperatures
of the studied Ni$_{50}$Mn$_{25+x}$Ga$_{25-x}$ alloys: forward
martensite transformation temperature $T_{M}$$\approx M_{S}\approx M_{F}$,
reverse martensite transformation temperature $T_{A}\approx A_{S}\approx A_{F}$,
forward IMT start temperature $T_{IMT}$, and reverse IMT start temperature
$T_{RIMT}$. Equilibrium temperature was calculated as $T_{0}=(T_{IMT}+T_{RIMT})/2$
for alloys 1--3, for Alloys 4 and 5 it was determined by extrapolation,
see Ref.~\cite{Straka_IMT}.}

\noindent {\small }\begin{tabular}{lcccccccc}
\multicolumn{1}{l}{} &  &  & \tabularnewline
\hline
\hline 
{\small Alloy} & x & {\small Nominal composition } & {\small Composition by XRFS } & {\small $T_{M}$ } & {\small $T_{A}$ } & {\small $T_{IMT}$ } & {\small $T_{RIMT}$ } & {\small $T_{0}$ }\tabularnewline
\hline 
 & {\small (at. \%)} & {\small (at. \%)} & {\small (at. \%)} & {\small (K)} & {\small (K)} & {\small (K)} & {\small (K)} & {\small (K) }\tabularnewline
\hline
\hline 
{\small Alloy 1} & 3.9 & {\small Ni$_{50.0}$Mn$_{28.9}$Ga$_{21.1}$} & {\small Ni$_{49.8}$Mn$_{29.4}$Ga$_{20.8}$} & {\small 328} & {\small 336} & {\small 251} & {\small 310} & {\small 281}\tabularnewline
{\small Alloy 2} & 3.7 & {\small Ni$_{50.0}$Mn$_{28.7}$Ga$_{21.3}$} & {\small Ni$_{50.2}$Mn$_{28.5}$Ga$_{21.3}$} & {\small 324} & {\small 330} & {\small 182} & {\small 287} & {\small 235}\tabularnewline
{\small Alloy 3} & 3.5 & {\small Ni$_{50.0}$Mn$_{28.5}$Ga$_{21.5}$} & {\small Ni$_{50.1}$Mn$_{28.4}$Ga$_{21.5}$} & {\small 318} & {\small 323} & {\small 85} & {\small 274} & {\small 178}\tabularnewline
{\small Alloy 4} & 3.2 & {\small Ni$_{50.0}$Mn$_{28.2}$Ga$_{21.8}$} & {\small Ni$_{50.0}$Mn$_{28.2}$Ga$_{21.8}$} & {\small 309} & {\small 315} & {\small 10} & \emph{\small not resolved} & {\small $\approx$100}\tabularnewline
{\small Alloy 5} & 2.7 & {\small Ni$_{50.0}$Mn$_{27.7}$Ga$_{22.3}$} & {\small Ni$_{50.0}$Mn$_{27.5}$Ga$_{22.5}$} & {\small 297} & {\small 301} & \multicolumn{2}{c}{\emph{\small no\,IMT\,above\,1.7\,K}} & {\small $\approx$0}\tabularnewline
\hline
\hline 
 &  &  &  &  &  &  &  & \tabularnewline
\end{tabular}
\end{table*}
 Five Ni$_{50}$Mn$_{25+x}$Ga$_{25-x}$ alloys for the study, where
$x$ was between 2.7 and 3.9 at.\%, Table~1, were produced by directional
solidification in Adaptamat Ltd. The alloys were essentially the same
as in our previous reports on the twinning stress \cite{Straka_IMT,Heczko1.7K}.
All alloys exhibited five-layered modulated (10M) martensite structure
at room temperature. This structure is approximated in this study
by a monoclinic lattice with the unit cell derived from the parent
cubic L2$_{1}$ cell \cite{Straka_Acta_2011}. Using the monoclinic
lattice allows to catch the main features of the structural changes
without getting entangled into complexity and details of still disputed
structure of 10M martensite. Limits of such approach are discussed
later in subsection~\ref{sub:3.5.Limits-of-the}.

The cuboid single crystal samples of dimensions of 1$\times$2.5$\times$10~mm$^{3}$
and 1$\times$2.5$\times$20~mm$^{3}$ were cut from heat treated
ingots along the \{100\} planes. All crystals exhibited MIR at room
temperature and very low twinning stress of $\approx$0.1 MPa for
Type 2 and $\approx$1 MPa for Type 1 twins. The temperature dependences
of twinning stress of alloys 1--5 were taken from Refs.~\cite{Straka_IMT}
and \cite{Heczko1.7K}, while the additional points for other alloys
with $x=$2.7--3.9 were taken from Ref. \cite{StrakaT1xT2}. 

The nominal compositions of the alloys and the compositions determined
using X-ray fluorescence (XRF) spectroscopy are given in Table~1
together with transformation temperatures. The main difference between
the alloys is their Mn/Ga content, represented by $x$. Keeping the
Ni content the same and as precisely as possible at 50~at.\% is critical
since the 10M phase region in Ni-content--temperature phase diagram
becomes narrow at low temperatures \cite{Straka_2011Jphys}. Even
a very small deviation of Ni content of the order of 0.1~at.\% may
result in enlarged twinning stress or instability of 10M martensite
(see supplementary material of Ref.~\cite{Straka_IMT}). The magnetic
and (inter)martensite transformation temperatures given in Table~1
were determined using AC and DC magnetic susceptibility measurements
of the particular studied samples, and by complementary optical observations
of twin bands (dis)appearance for the case of (reverse) martensite
transformation.

The XRD measurements on single crystals were performed using two laboratory
diffractometers with parallel beam optics and Euler cradle. We had
to resort to non-usual X-ray analysis of single crystal in order to
study precisely the same single crystals which exhibited the very
low twinning stress and MIR. In previous study Mogylnyy et al. \cite{Mogylny}
demonstrated that on single crystals of 10M martensite the slight
lattice monoclinicity can be seen well as the separation of the relevant
diffraction lines such as (400) and (040), and (440) and ($\bar{4}$40)
(adapted to our notation, originally (2~0~10) and (2~0~$\bar{10}$),
and (200) and (0~0~10)). The (400), (040) and (004) diffraction
lines were measured in Bruker D8 Discover diffractometer equipped
with rotating Cu anode ($\lambda=0.1540598$~nm) and cooling stage
Anton Paar DCS 350. The stage temperature was varied from 350~K to
170~K. The (600), (060), (440), and ($\bar{4}$40) diffraction lines
were measured in PANalytical X'Pert Pro diffractometer equipped with
Co anode ($\lambda=0.178901$~nm) and in-house built heating/cooling
stage based on Peltier element. The superstructure \{600\} diffraction
lines offer more precise lattice parameter determination than \{400\}
diffraction lines, but at the cost of small diffracted intensity ($\approx$200
times lower than for \{400\}) \cite{Straka_Acta_2011}. In addition
to limited amount of the lines, the precision of the structural parameters
was limited by broadening of the martensitic peaks. The width of peaks
was at least 0.2$^{\circ}$ compared to 0.09$^{\circ}$ for laboratory
standard of Si single crystal. 

To get unambiguous and as precise as possible lattice parameters we
prepared samples with uniform orientation of c-axis ({}``single variant''
state) by a few MPa compression, i.e., neither Type 1 nor Type 2 twin
boundary was present during the XRD measurements. Nonetheless, the
sample with this uniform orientation of c-axis still exhibits rich
internal structure. It typically contains internal \{100\} compound
twins and internal \{110\} compound twins, referred also as $a/b$-laminate
and modulation domains, respectively \cite{Straka_Acta_2011,Chulist_ab_laminate}.
The unavoidable presence of the $a/b$-laminate allows to observe
the (400) and (040) diffraction lines for single orientation of the
sample; same applies also for the (600), (060) or (440), ($\bar{4}$40)
pairs. 

The diffraction maxima of the single crystals were first located using
$\omega$- and $\psi$- scans. Then the $\omega-2\theta$ scans were
measured with corresponding offsets. The obtained diffractograms were
evaluated by in-house software that fitted up to six peaks using Pearson
VII functions \cite{Prevey}. To achieve relevant precision, the peaks
were fitted using $K_{\alpha}$ doublet. The width and shape parameters
of Pearson VII function were constrained to have the same value for
one\textbf{ }diffractogram. That gave good stability of the fit when
diffraction lines were overlapped at the cost of slightly reduced
fit precision as the assumption of the same width for all diffraction
lines was not fully justified.

In order to determine lattice parameter $\gamma$, we measured the
\{440\} diffraction lines as they are significantly influenced by
this angle. In the monoclinic structure, the equation for the \{$hkl$\}
diffraction lines is \cite{Book_XRD}:

\begin{equation}
\frac{1}{d_{hkl}^{2}}=\dfrac{\dfrac{h^{2}}{a^{2}}+\dfrac{k^{2}}{b^{2}}-\dfrac{2hk\cos\gamma}{ab}}{sin^{2}\gamma}+\dfrac{l^{2}}{c^{2}}.\label{eq:gammaXRD}\end{equation}

The two (440) and ($\bar{4}$40) diffraction lines in combination
with (400) and (040) lines -- or for increased precision (600) and
(060) lines -- provided all necessary information for $\gamma$ determination.
We had four independent measurements to determine three parameters:
a, b, and $\gamma$. The interplanar distance $d_{hkl}$ was calculated
using Bragg's law $2d_{hkl}\sin\theta=n\lambda$. The search for \{440\}
diffraction lines, however, turned to be somewhat laborious when using
powder diffractometers in single crystal studies. Therefore we developed
a complementary method for $\gamma$ determination, which utilized
the fact that $\gamma$ angle is closely related to the angle $\alpha$
observed between the traces of Type 1 and Type 2 twin boundaries on
\{100\} oriented surface \cite{Sozinov_Type2} (see also Fig.~\ref{fig:gamma_determination}):
\begin{equation}
\cos\gamma=\frac{c^{2}-b^{2}}{2ab}\tan\alpha.\label{eq:gammaOptical}\end{equation}

It is important to note here that even very small monoclinic distortion
$(\gamma-90^{\circ})$ of the order of $0.1^{\circ}$ can result in
relatively large angle $\alpha$ of the order of several degrees observed
optically on the surface \cite{Sozinov_Type2,StrakaT1xT2,Straka_Acta_2011}.
The temperature dependences of lattice parameters $a(T)$, $b(T)$,
and $c(T)$ were determined from \{400\} diffraction lines. The $\alpha(T)$
dependence was obtained from optical observations of sample with both
Type 1 and Type 2 twin boundaries close to each other, using a light
microscope equipped by an in-house built cooling/heating stage. Equation~\ref{eq:gammaOptical}
is, however, valid only for ideal \{101\} twins without internal structure.
We assumed regular $a/b$-lamination, i.e. the same volume fraction
of $a$- and $b$-oriented lamellas ($\lambda=0.5$ according to notation
of Ref.~\cite{Straka_Acta_2011}), and used a relevantly modified
equation:

\begin{equation}
\gamma=\frac{1}{2}\arccos(\frac{c^{2}-b^{2}}{2ab}\tan\alpha)+\frac{1}{2}\arccos(\frac{c^{2}-a^{2}}{2ab}\tan\alpha)\label{eq:GammaOpticalLaminate}\end{equation}

Fine modulation domains can also lead to various tilt of Type 2 twin
boundary and a false $\alpha$ reading \cite{Heczko_Acta2013,Heczko_MRB}.
Nonetheless, in contrast to $a/b$-laminate, the modulation domains
are often large enough (at least for crystals from Adaptamat) to be
identified in optical microscope \cite{Straka_Acta_2011} and are
also more easily controlled, for example by mechanical training \cite{Chulist_modulation}.
We avoided the effect of modulation domains by preferably selecting
samples with very large or nearly single modulation domain. In some
cases, mechanical training consisting of tensile/compressive loadings
was used to change the distribution of modulation domains towards
the single domain configuration.

\section{Results and discussion}

In the following subsections \ref{sub:3.1.The-10M14MNM-transformation}
and \ref{sub:3.2.Temperature-dependence-of} we describe in detail
the study of two alloys (alloy 1 and 3) representing typical behavior
and then we summarize all observations for all five alloys in subsections
\ref{sub:3.3Temperature-dependence-of-1} and \ref{sub:3.4.Temperature-dependence-of}.
In subsection \ref{sub:3.5.Limits-of-the} we discuss the limits of
the used lattice approximation. The last two subsections \ref{sub:3.6.Relation-betwen-lattice}
and \ref{sub:3.7.Relation-betwen-lattice} provide the comparison
of structure evolution with the measured twinning stress. The first
subsection \ref{sub:3.1.The-10M14MNM-transformation} deals with simple
case on which the validity of the structure determination method is
demonstrated.

\subsection{\label{sub:3.1.The-10M14MNM-transformation}The 10M$\leftrightarrow$14M$\leftrightarrow$NM
transformation sequence observed in alloy 1}

\begin{figure}[!t]
\includegraphics[width=77mm]{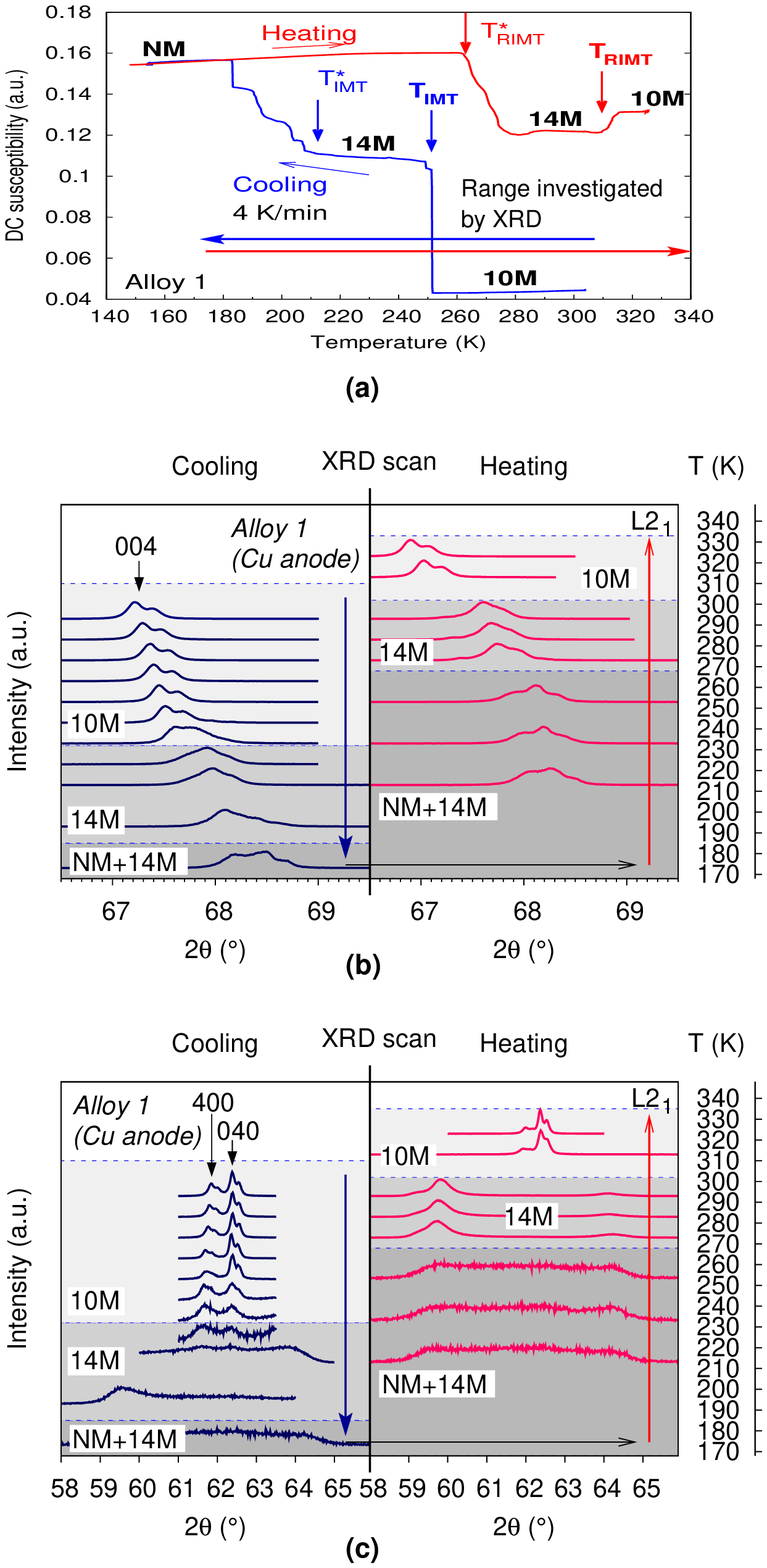}

\caption{\label{fig:c_alloy1} Structural changes in alloy 1: a) DC magnetic
susceptibility curve with intermartensite transformation temperatures
and corresponding phases 10M, 14M, NM marked. b, c) $\omega-2\theta$
scans in selected $2\theta$ intervals performed during quasistatic
cooling and heating in temperature range marked in (a). The patterns
are normalized to maximum intensity and positioned according to the
measurement temperature (axis on the right). The \{400\} peaks of
10M phase and regions with different phases 10M, 14M, NM are marked.
Peak splitting due to $K_{\alpha}$ doublet is marked in Fig.~\ref{fig:c_alloy3}b. }

\end{figure}
The magnetic susceptibility measured for alloy 1 during cooling and
subsequent heating is shown in Fig.~\ref{fig:c_alloy1}a. During
cooling from 310~K, there are no significant changes in susceptibility
down to $T_{IMT}=$251~K, where a large sharp jump starts. This first
jump in susceptibility is ascribed to the transformation to 14M martensite.
During further cooling, start of second jump occurs at $T_{IMT}^{*}$,
which marks the transformation of the 14M martensite to so-called
non-modulated (NM, purely tetragonal) martensite with long $c$-axis.
Upon following heating, the material exhibits again two sharp changes
in susceptibility, ascribed to the reverse transformations NM$\rightarrow$14M
and 14M$\rightarrow$10M at $T_{RIMT}^{*}$ and $T_{RIMT}$, respectively.
The different initial and final DC magnetic susceptibility for 10M
martensite is easily explained by different twin variant distributions
before and after the transformations to other phases.

The 10M$\leftrightarrow$14M$\leftrightarrow$NM intermartensite transformation
(IMT) sequence is well known and was presented previously e.g. in
\cite{segui2005,segui2007,Chernenko_Stress_temp}. As the temperature
range investigated by XRD includes the IMTs of alloy 1, Fig.~\ref{fig:c_alloy1}a,
it is obvious that all the mentioned IMTs shall be reflected in the
XRD patterns.

The thermal evolution of (004) peak in $\omega-2\theta$ scans performed
during cooling and heating is displayed in Fig.~ \ref{fig:c_alloy1}b.
This and all below discussed diffraction peaks are split due to the
presence of $K_{\alpha}$ doublet in the diffraction spectrum. During
cooling from room temperature, the (004) peak shifts gradually, indicating
the gradual shortening of the $c$ lattice parameter. At 230~K, the
peak broadens and then it shifts suddenly to the right at 220~K.
That indicates 10M$\rightarrow$14M transformation with the mixture
of two phases being temporarily present around 230~K. The sudden
shift to the larger $2\theta$ marks the sudden contraction of the
$c$ lattice parameter and the finish of 10M$\rightarrow$14M IMT
(i.e., $c_{10M}$ changed to $c_{14M}$ and $c_{10M}>c_{14M}$). 

Further cooling resulted in another change of the (004) peak shape
at 170~K, at which temperature the peak consisted of two convoluted
lines (not counting the $K_{\alpha}$ split). The new line at $2\theta\approx68.5^{\circ}$
can be ascribed to NM phase; so there is a mixture of NM and 14M martensite
at 170~K. Similarly as for the 10M$\rightarrow$14M transformation,
the NM phase exhibited slightly shorter lattice parameter than the
14M phase. Note that for NM martensite, the described {}``(004)''
line actually corresponds to the (400)$_{NM}$ line and to the short
$a_{NM}$ lattice parameter (not to $c_{NM}$ parameter). The different
lattice parameters $c_{10M}>c_{14M}>a_{NM}$ indicate that the 10M$\rightarrow$14M$\rightarrow$NM
sequence can be induced also by an external compressive stress $\sigma_{EXT}$,
since the stress will preffer the shorter lattice parameter of the
other phase \cite{Chernenko_Stress_temp}. 

The observed transformation to NM martensite is, however, clearly
incomplete. The magnetic susceptibility curve indicates that the whole
14M$\rightarrow$NM transformation occurs in about 30~K interval
and sharply ends, Fig.~\ref{fig:c_alloy1}a. Additional cooling beyond
the limit of our experimental arrangement would presumably result
in a pure NM phase. During heating from 170 K, the reverse transformations
can be seen in the XRD pattern as the sudden shifts of (004) peak
towards smaller $2\theta$, Fig.~\ref{fig:c_alloy1}b. These shifts
correspond to the reverse transformation sequence NM(+14M)\-$\rightarrow$\-14M\-$\rightarrow$\-10M,
and to corresponding reverse changes of the relevant lattice parameter.
The structural changes 10M$\leftrightarrow$14M$\leftrightarrow$NM
during cooling and heating are thus clearly demonstrated by the changes
of the lattice parameter corresponding to the {}``(004)'' peak,
Fig.~\ref{fig:c_alloy1}b.

The thermally-induced structural changes in alloy 1 are even more
visible when monitoring the (400) and (040) diffraction lines, i.e.
$a$ and $b$ lattice parameters of 10M martensite. The two reflections
shift slightly with the decreasing temperature indicating gradual
changes in $a$ and $b$ lattice parameters, but they suddenly disappear
at about 230~K, Fig.~\ref{fig:c_alloy1}c. Instead of these two
reflections, two other lines appear at $2\theta\approx59.5^{\circ}$
and at $2\theta\approx64^{\circ}$. That corresponds very well to
the (400)$_{14M}$ and (040)$_{14M}$ reflections, previously reported
in the literature \cite{Sozinov_14M,Cakr_IMT_2013,Kaufman_Adaptive_PRL},
and thus we can be quite confident that we really observe the 14M
phase. Upon further cooling, these peaks almost disappear at 180~K
due to the transformation to NM martensite. Upon following heating,
the corresponding reverse transformations occur, resulting in reappearance
of the relevant peaks, Fig.~\ref{fig:c_alloy1}c.

In summary, we can conclude that alloy 1 exhibits behavior which is
expected from the previously known 10M$\leftrightarrow$14M$\leftrightarrow$NM
transformation sequence. The changes in \{400\} lines of 10M martensite
or corresponding lines of the other phases reflect the thermally-induced
changes in 10M lattice and also clearly indicate the IMTs of the 10M$\leftrightarrow$14M$\leftrightarrow$NM
sequence. The determined lattice parameters of the individual phases
have relation $c_{10M}>c_{14M}>a_{NM}$. Importantly we observed also
a mixture of 10M+14M and 14M+NM martensites, however, they were only
present in limited temperature intervals. The confirmed behavior gave
us the confidence that the used method is sound and can be applied
to more complicated cases as shown below.

\subsection{\label{sub:3.2.Temperature-dependence-of}Temperature dependence
of $a,\, b,\, c$ lattice parameters in alloy 3}

\begin{figure}[!t]
\includegraphics[width=77mm]{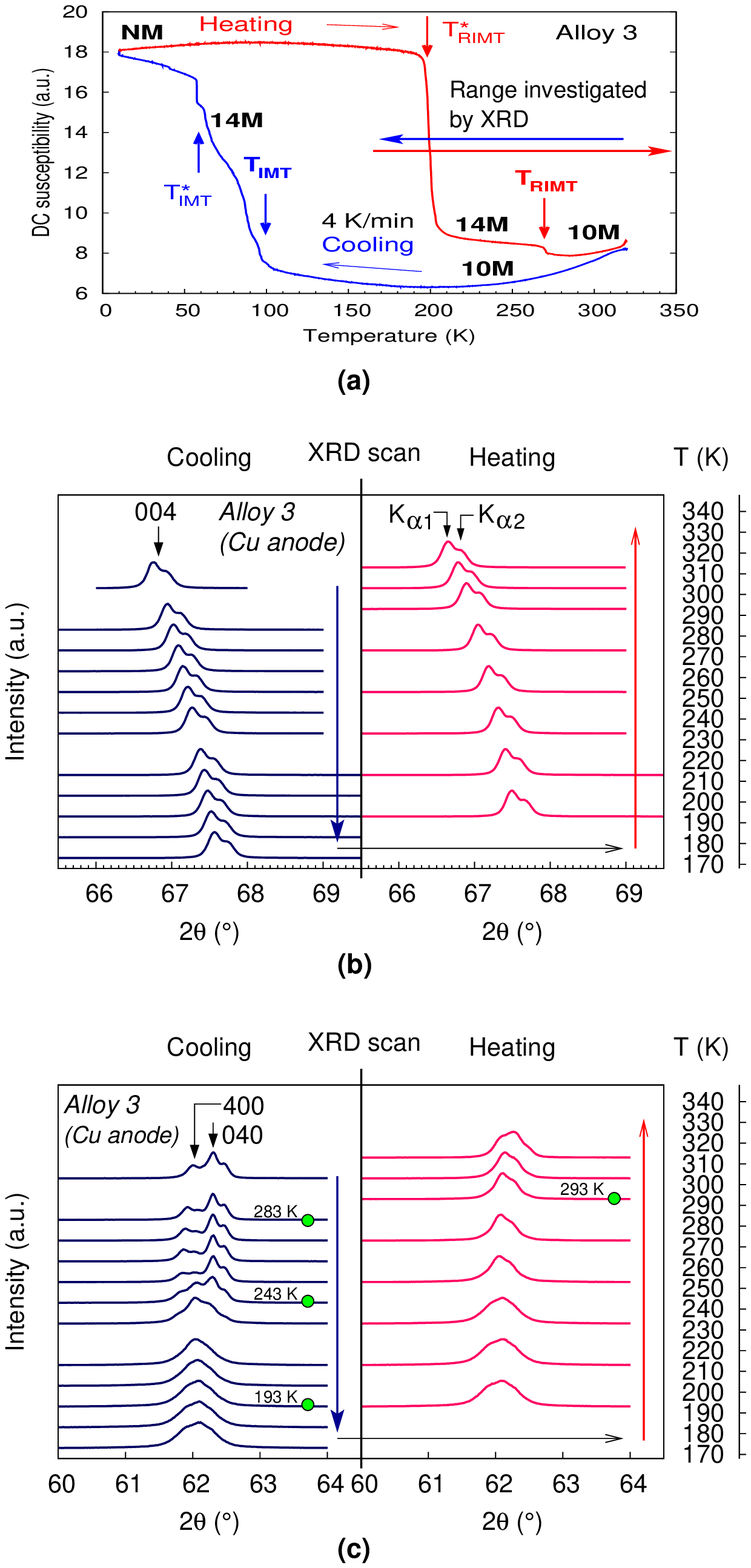}

\caption{\label{fig:c_alloy3} Structural changes in alloy 3: a) DC magnetic
susceptibility with ascribed intermartensite transformation temperatures
and corresponding phases 10M, 14M, NM marked. b, c) $\omega-2\theta$
scans in selected 2$\theta$ intervals performed during quasistatic
cooling and heating in temperature range marked in (a). The patterns
are normalized to maximum intensity and are positioned according to
the measurement temperature (axis on the right). The \{400\} peaks
of 10M phase and peak splitting due to $K_{\alpha}$ doublet are marked.}

\end{figure}
The magnetic susceptibility measured for alloy 3 during cooling and
following heating is shown in Fig.~\ref{fig:c_alloy3}a. The susceptibility
curve exhibits similar features as the curve for alloy 1 indicating
the 10M$\leftrightarrow$14M$\leftrightarrow$NM transformation sequence,
but the transformations are shifted to much lower temperature and
are less clearly separated. The intermartensite transformation temperatures
$T_{IMT}$ and $T_{IMT}^{*}$ are well below the interval available
in the X-ray diffraction measurement and thus none of the 10M$\leftrightarrow$14M$\leftrightarrow$NM
IMTs can be seen in the XRD patterns.

The (004) peak for alloy 3 monitored during cooling and heating is
shown in Fig.~\ref{fig:c_alloy3}b. The peak gradually shifts with
temperature indicating the gradual changes in $c$ lattice parameter,
but there are no sudden shifts as those observed for alloy 1. That
is an additional indication that none of the 10M$\leftrightarrow$14M$\leftrightarrow$NM
transformations occurs. Nonetheless, some subtle changes in structure
appear, reflected as changes in (400) and (040) peaks, described below.
During cooling, the (400) and (040) peaks only shift slightly with
the decreasing temperature at first, Fig.~\ref{fig:c_alloy3}c. At
243\,K, the peaks suddenly start changing their shape, and at even
lower temperature, the two peaks (400) and (040) merge into a single
broad peak which looks almost featureless. During the following heating,
this broad peak changes only slightly its shape but does not visibly
split.

Closer analysis of the selected XRD patterns obtained at 283, 243,
and 193\,K upon cooling and at 293\,K upon heating (marked in Fig.\,\ref{fig:c_alloy3}c
by filled green circles) is shown in Fig.~\ref{fig:peak_analysis}.
The analysis reveals that in addition to the two (400) and (040) lines
observed e.g. at 283\,K, Fig.~\ref{fig:peak_analysis}a, a third
line appears around 243\,K, Fig.~\ref{fig:peak_analysis}b. With
the temperature decreasing further, this new peak gains intensity
on the account of the original (400) peak, Fig.~\ref{fig:peak_analysis}c.
We assign $a'$ lattice parameter to this new line, where $a>a'>b$.
As we monitor only few peaks, we cannot decide here whether the new
line reflects the growth of {}``new'' martensitic phase or if the
same lattice is showing a new type of distortion. The detailed analysis
using synchrotron radiation is planned to clarify the issue. Upon
following heating from low temperatures, the peak shape also changes
with temperature, and the analysis indicates that at 273\,K, the
XRD pattern can be fit by only a single peak, corresponding to a common
lattice constant $a=b$. 

Thus, we observe some kind of structural transformation which results
in sudden small sharp changes of $a$ and $b$ lattice parameters
but importantly not of $c$ parameter. Similar XRD pattern developments,
corresponding to sudden sharp changes in $a$ and $b$ or to $a=b$,
were observed also in alloys 4 and 5. In these cases, however, no
third peak was found. All observations are summarized and discussed
in the next chapter.

\subsection{\label{sub:3.3Temperature-dependence-of-1}Temperature dependence
of $a$,\,$b$,\,$c$ lattice parameters summarized for all alloys}

\begin{figure*}
\includegraphics[width=15cm]{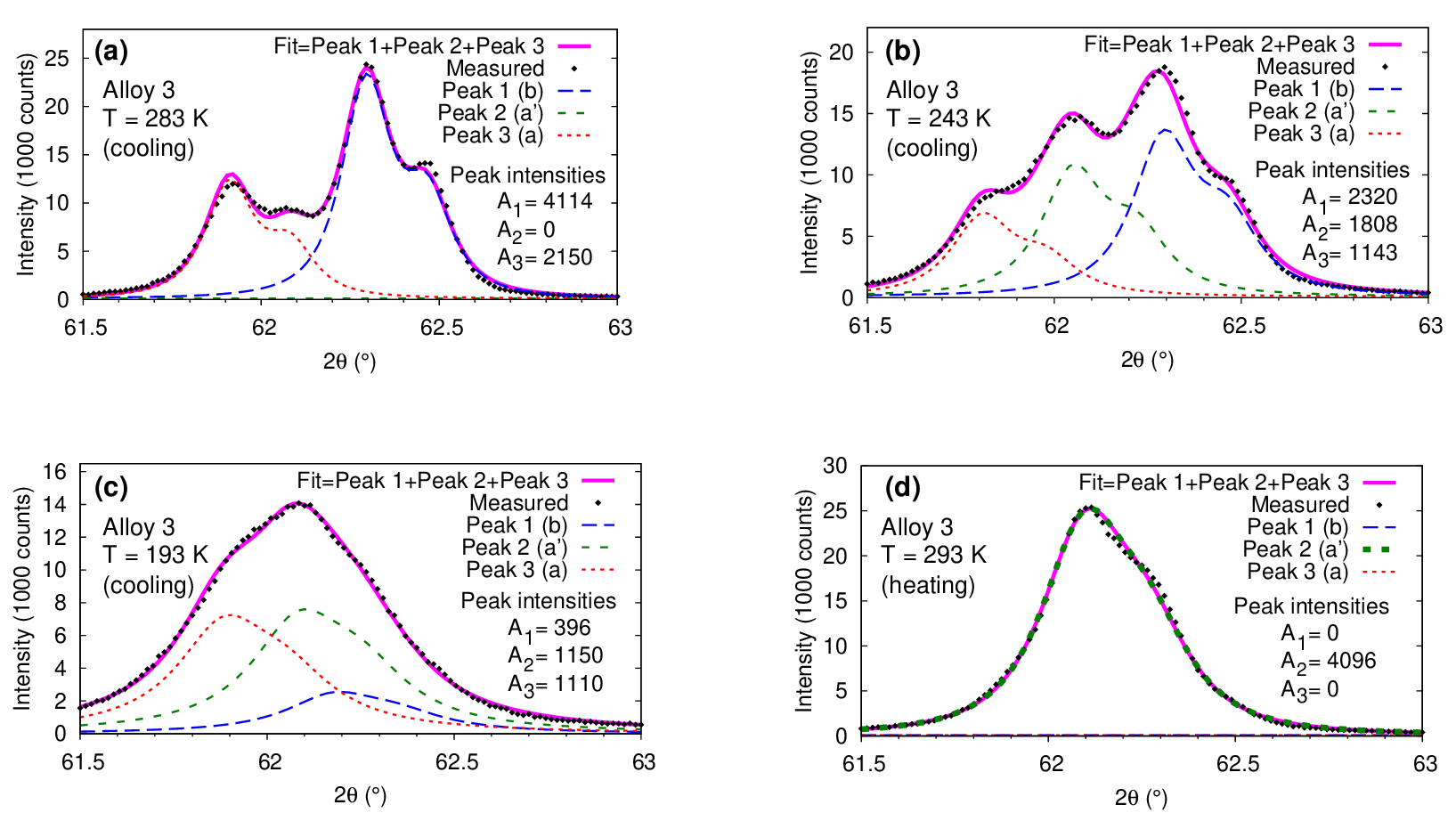}\caption{\label{fig:peak_analysis}\emph{ }The closer look and fitting of selected
XRD patterns (marked by green filled circles in Fig.~\ref{fig:c_alloy3})
using Pearson VII function: a) alloy 3 at 283 K, cooling; b) alloy
3 at 243 K, cooling; c) alloy 3 at 193\,K, cooling; d) alloy 3 at
293 K, heating. The provided peak intensities are normalized to Lorentz
polarization factor ($\approx$5).}

\end{figure*}
\begin{figure*}
\includegraphics[width=16cm]{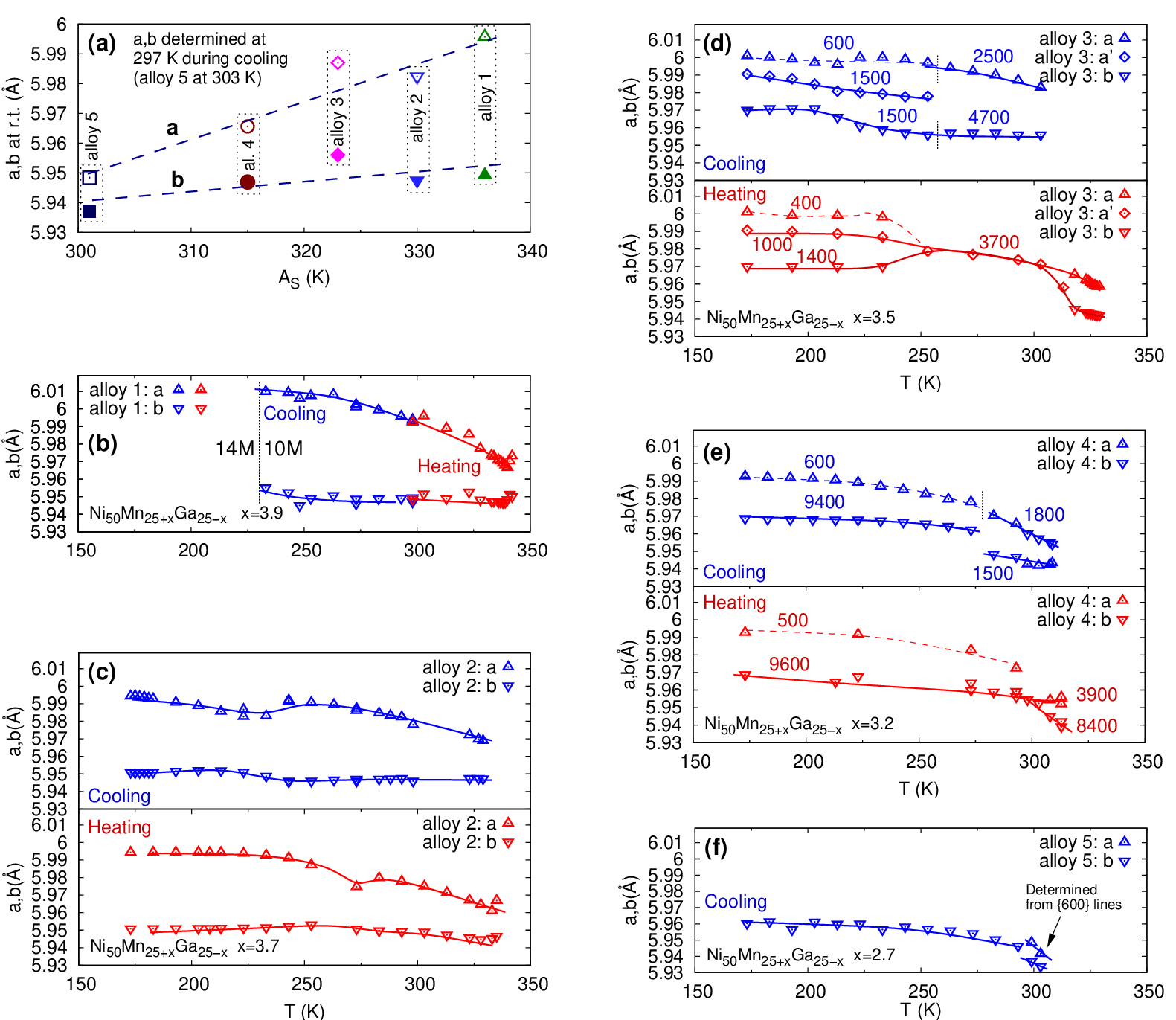}\caption{\label{fig:a_b_summary} Temperature dependences of $a,b$ lattice
parameters summarized for alloys 1--5: a) Room temperature $a,b$
lattice parameters as functions of austenite start ($A_{S}$) temperature.
b-f) Determined temperature dependences of $a,b$ lattice parameters
for alloy 1 (b), alloy 2 (c), alloy 3 (d), alloy 4 (e), and alloy
5 (f). The numbers next to selected curves are average peak intensities
(counts) of the relevant curve parts. Dashed and solid lines are just
guides for eyes.}

\end{figure*}
The room temperature $a,b$ lattice constants determined during cooling
are summarized in Fig.~\ref{fig:a_b_summary}a. In agreement with
the previous investigation by Lanska et al. \cite{Lanska_2004}, the
difference between $a$ and $b$ decreases when the (reverse) martensite
transformation temperature approaches the room temperature.

The $a$ and $b$ lattice parameters of all alloys as functions of
temperature are displayed in Fig.\,\ref{fig:a_b_summary}b-f. Alloy
1 exhibits small gradual changes of the parameters with temperature,
Fig.\,\ref{fig:a_b_summary}b, with $b$ almost constant and $a$
rising slightly with decreasing temperature. Pagounis et al. reported
recently same trends in the lattice constants for Ni$_{50}$Mn$_{29.2}$Ga$_{20.8}$,
which is very close to alloy 1 \cite{Pagounis_abc}. Alloy 2 exhibits
similar dependence, but the parameters show some tendency to come
closer to each other at about 220~K upon cooling and at about 270~K
upon heating, Fig.\,\ref{fig:a_b_summary}c. In alloy 4, the parameters
seem to actually coincide at about 270~K upon cooling and separate
at about 300~K upon heating, but there is also a weak line corresponding
to the original (400) line or $a$ parameter, with decreasing intensity,
Fig.\,\ref{fig:a_b_summary}e. Similar coincidence of parameters
upon cooling is observed in alloy 5, in which, however, no weak line
is observed and the parameters are so close to each other that they
can be distinguished only by using \{600\} diffraction lines, which
provide better resolution than \{400\}, Fig.\,\ref{fig:a_b_summary}f.
Unfortunately the analysis using \{600\} lines was only possible near
room temperature in our experimental arrangement. 

Alloy 3 exhibits complex development of lattice constants, Fig.\,\ref{fig:a_b_summary}d,
which may be a combination of the effects observed in alloys 2 and
4. Upon cooling, the parameters come closer to each other at about
260~K (but do not coincide) while there is still an extra weak line
corresponding to the original (400) reflection. During heating, the
parameters eventually coincide at 250~K and then separate around
320~K (see also peak analysis in Fig.~\ref{fig:peak_analysis}).
In repeated experiments, the weak lines were sometimes undetected
in alloys 3 and 4, which may be due to different analyzed spot or
sample adjustment.

\begin{figure}[tb]
\includegraphics[width=77mm]{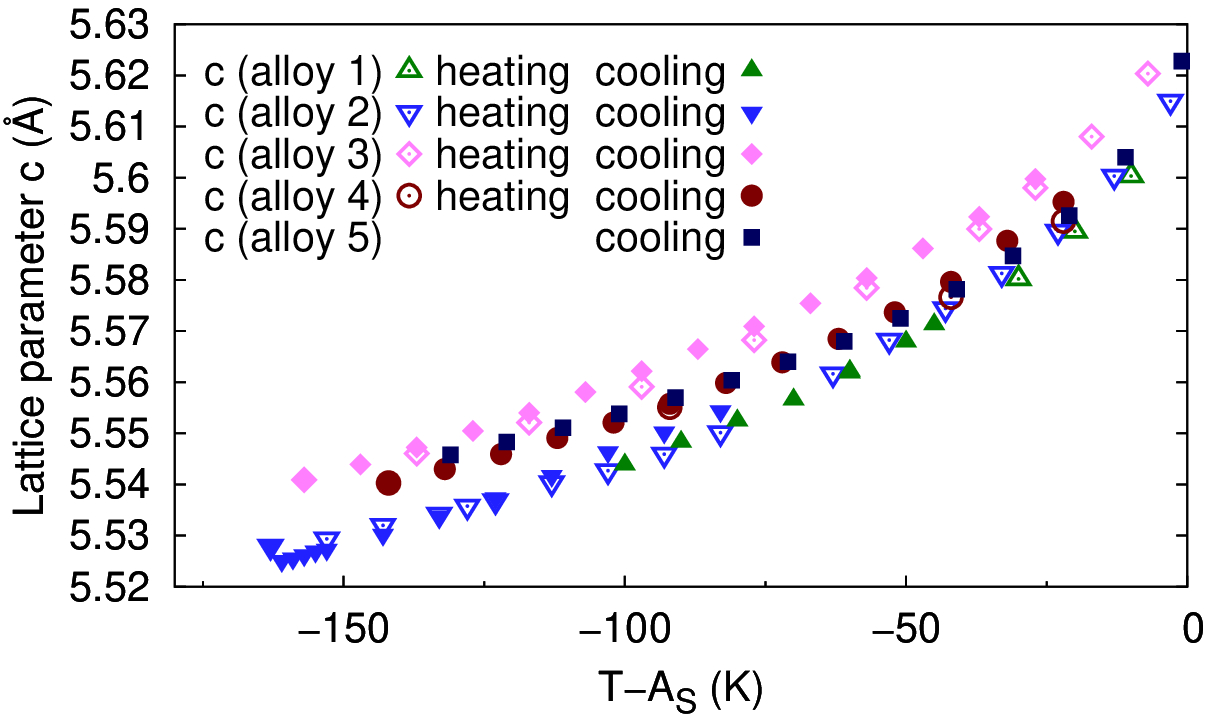}\caption{\label{fig:c_constant} Lattice parameter $c$ as a function of relative
temperature $(T-A_{S})$ for alloys 1--5. ~~~~~~~~~~~~~~~~~~~~~~~~~~~~~~~~~~~~~}

\end{figure}

In contrast to complex changes observed for $a,b$ lattice parameters,
the $c$ lattice parameter exhibits rather uniform behavior in all
alloys. The dependence of $c$ lattice parameter of 10M martensite
on the relative temperature $(T-A_{S})$ is similar in all alloys
studied, Fig.~\ref{fig:c_constant}; the parameter decreases gradually
with decreasing temperature. 

\begin{figure}[tb]
\includegraphics[width=77mm]{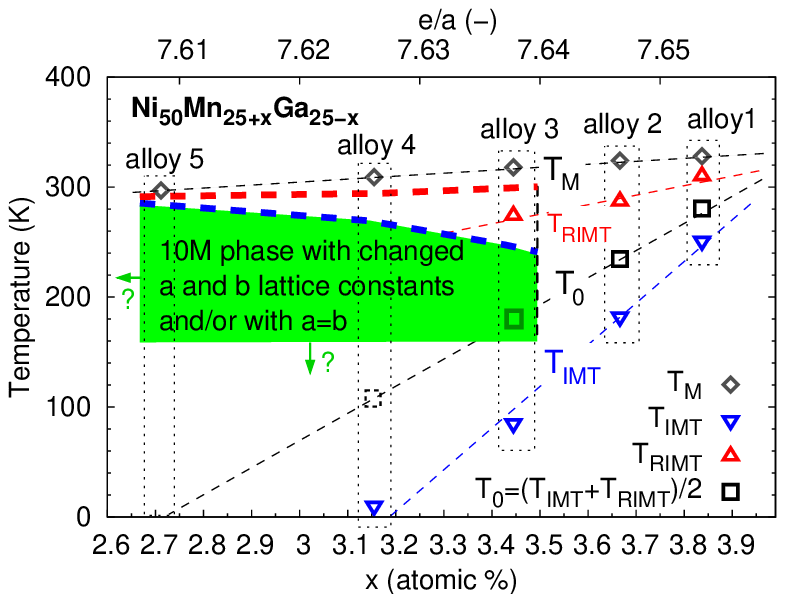}\caption{\label{fig:phase_diagram} Phase diagram showing the region with {}``new
phase'' with changed $a,b$ lattice parameters and/or with $a=b$.
The region is marked by green color and thick dashed blue line (changes
in $a,b$ occurring on cooling) and red line (changes in $a,b$ occurring
on heating).}

\end{figure}

Based on $a,b,c$ lattice parameters evolution obtained from the peak
analysis, we can state rather confidently that some significant changes
in 10M structure related only to $a$ and $b$ lattice parameters
occur in alloys 3 and 4 upon heating and cooling. The approximate
temperature and compositional region of this {}``new phase'' and
of phase with $a=b$ is marked by the green area in the phase diagram
in Fig.~\ref{fig:phase_diagram}. New phases were reported in Ni-Mn-Ga
before; for example Kim et al. \cite{Kim_X_Phase} and Kushida et
al. \cite{Kushida_X_Phase} indicated new {}``x-phase'' induced
in austenite or pre-martensite by compressive stress. However, as
we investigate only few lines of a single crystal diffraction pattern,
we cannot provide full explanation of the new structure formed. That
is beyond the scope of this article and requires further research.
Here we can only suggest that for certain composition and temperature
ranges, the material transforms to a slightly modified or {}``new''
10M phase. In our monoclinic approximation this phase exhibits $a$
close to or it is even identical to $b$ (corresponding to the strong
$a'$ or $b$ lines at low temperatures in Fig.~\ref{fig:a_b_summary}d-f),
while the residua of the original phase with $a\neq b$ remain in
the material (and generate the weak (400) or $a$ line). See section~\ref{sub:3.5.Limits-of-the}
for further discussion.

\subsection{\label{sub:3.4.Temperature-dependence-of}Temperature dependence
of $\gamma$ lattice parameter}

\begin{figure}[!tbh]
\includegraphics[width=77mm]{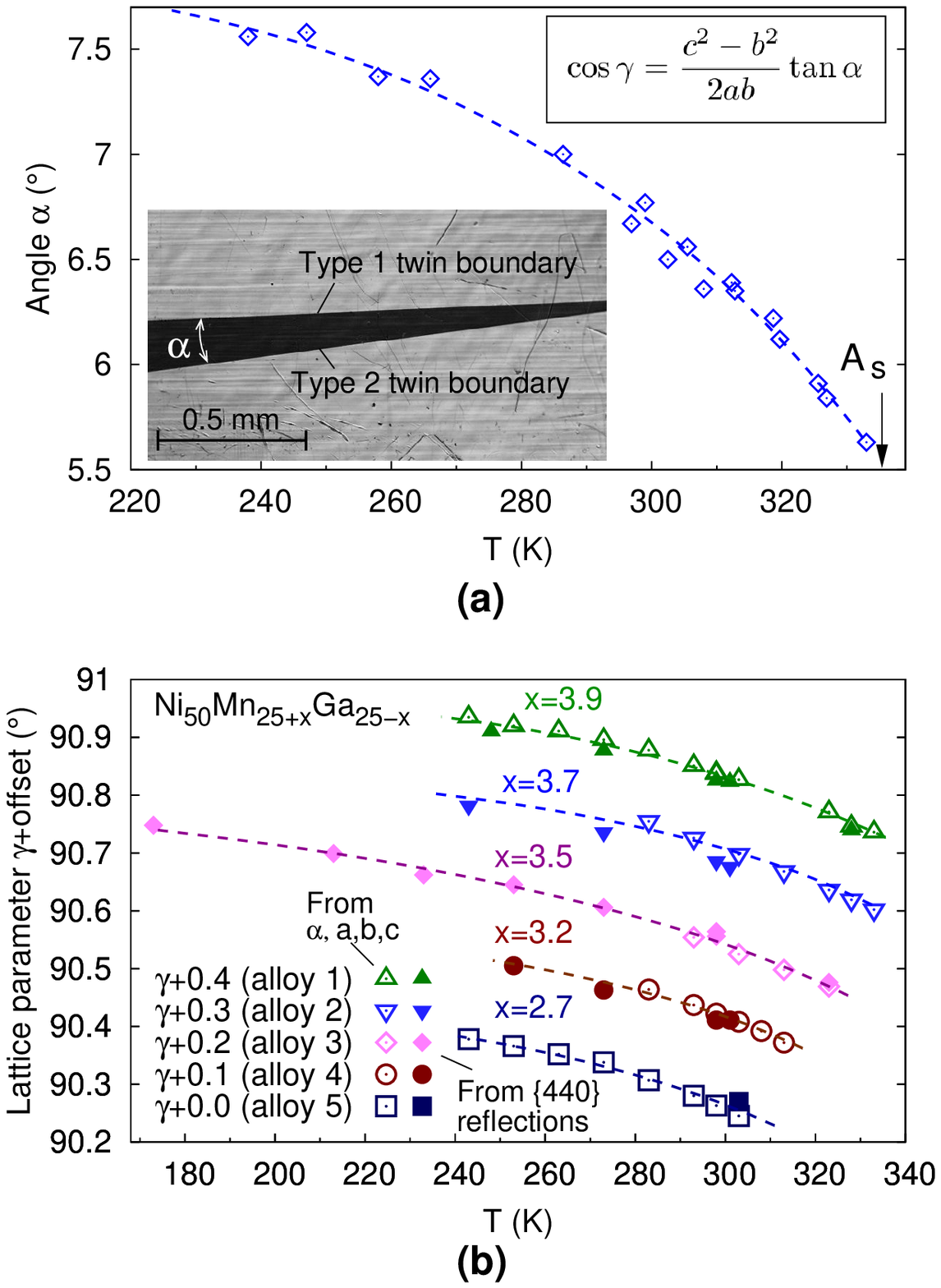}\caption{\label{fig:gamma_determination}Measurements related to $\gamma$
lattice parameter: a) Angle $\alpha$ as a function of temperature
in alloy 1, determined as illustrated in the lower inset. The principal
relation between $\alpha$ and $\gamma$ is given in the upper inset.
b) Lattice parameter $\gamma$ (\emph{+offset}) as a function of temperature
determined from \{440\} reflections, Eq.~\ref{eq:gammaXRD} (filled
symbols) and from $\alpha$, Eq.~\ref{eq:GammaOpticalLaminate} (empty
symbols) for alloys 1--5. Note various 0--0.4$^{\circ}$ offset added
to the dependences for the sake of clarity, for an alternative offset-free
plot, see Fig.~\ref{fig:comparison_TS}c.}
 
\end{figure}
The $\gamma$ lattice parameter was determined by two methods: from
\{440\} reflections (Eq.~\ref{eq:gammaXRD}) and from optical observations
of the angle $\alpha$ between the Type 1 and Type 2 twin boundary
traces on the \{100\} surface (insets in Fig.~\ref{fig:gamma_determination}a
and Eq.~\ref{eq:GammaOpticalLaminate}). The evident change of angle
$\alpha$ with temperature is demonstrated for alloy 1 in Fig.~\ref{fig:gamma_determination}a.
The $\alpha$ angle decreases with increasing temperature in all alloys.
However, even very near (reverse) martensite transformation it is
far from zero in all alloys, indicating that $\gamma$ deviates from
90$^{\circ}$ even just prior to reverse transformation. 

To demonstrate equivalence between two approaches the comparison is
made in Fig.~\ref{fig:gamma_determination}b; the filled symbols
were determined using Eq.~\ref{eq:gammaXRD}, while the open symbols
by Eq.~\ref{eq:GammaOpticalLaminate}. It is apparent that both methods
yield very similar values of $\gamma$. In order to facilitate the
comparison with the twinning stress, an alternative plot of $\gamma$
as a function of relative temperature $T-A_{S}$ is shown in Fig.~\ref{fig:comparison_TS}c.
All alloys exhibit very similar $\gamma(T-A_{S})$ dependence with
$\gamma$ decreasing with increasing temperature. Near martensite
transformation, $\gamma\approx90.25^{\circ}$, while 50~K below the
transformation, $\gamma\approx90.4^{\circ}$.

\subsection{\label{sub:3.5.Limits-of-the}Limits of the used lattice approximation}

The used monoclinic lattice approximation and description by $a,b,c,\gamma$
lattice constants cannot in principle describe fully the 10M structure
and its fine structural changes. The changes in diffraction patterns
observed in monoclinic approximation as sudden changes in $a,b$ lattice
constants may originate also from other effects than the simple change
in lattice symmetry. These may be, for example, refinement in the
$a/b$-lamination, changes in twinning periodicity, changes in stacking
of basal planes of 10M structure, or, more generally, as refining
or coarsening of adaptive martensite \cite{Kaufman_Modulated_Mart,Kaufman_Adaptive_PRL}.\emph{
}Recently Ge et al. \cite{Ge_coarsening} demonstrated gradual change
of lattice parameters resulting from the coarsening of nanotwins during
the 14M$\rightarrow$NM transformation observed by TEM. All the mentioned
effects can significantly influence the diffraction pattern and can
result in an additional or missing diffraction peaks and consequent
difficulties in lattice symmetry determination \cite{Ustinov,Ustinov_2}. 

In this respect it is also interesting to note that according to Righi
et al. \cite{Righi_MSF}, the transformation of the 10M structure
from commensurate to incommensurate did not result in sudden changes
in $a,b$ lattice constants. Additionally Glavatskyy \cite{Glavatsky_magnetic}
reported magnetic transitions in the 10M structure, but did not find
any sudden changes in lattice constants. In our case, the observed
structural transitions do not seem to be of magnetic character, since
we did not detect any significant changes in magnetic susceptibility
during the sudden small changes in $a,b$ lattice parameters (compare
Fig.~\ref{fig:c_alloy3}a and Fig.~\ref{fig:c_alloy3}c at 240~K).

\subsection{\label{sub:3.6.Relation-betwen-lattice}Relation between lattice
parameters and twinning stress for Type 1 twins}

\begin{figure*}
\includegraphics[width=15cm]{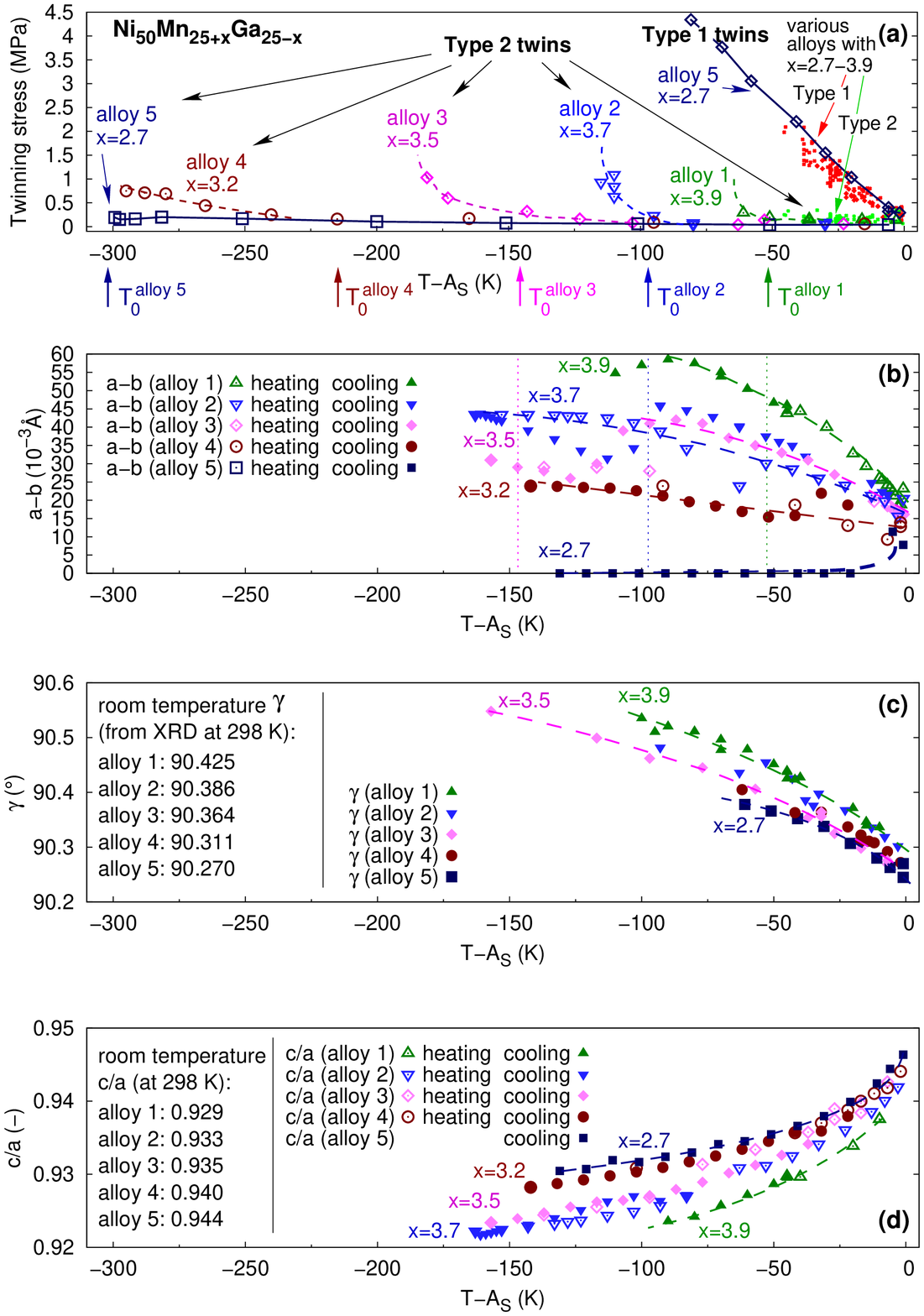}\caption{\label{fig:comparison_TS} Comparison of twinning stress temperature
dependences (a) with the temperature dependences of lattice parameters
$(a-b)$ (b), $\gamma$ (c), and $c/a$ (d). The temperature is given
relative to the austenite start temperature ($A_{S}$); $T_{0}=(T_{IMT}+T_{RIMT})/2$
marks the equilibrium temperature. The twinning stress dependences
are compiled from our previous measurements presented in Refs. \cite{Straka_IMT}
(alloy 1--5, Type 2 twins), \cite{Heczko1.7K} (alloy 5, Type 1 twins),
and \cite{StrakaT1xT2} (Type 1 and Type 2 twins, red and green filled
squares). Room temperature $\gamma$ and $c/a$ are additionally listed
in insets in (c) and (d). Dashed and solid lines are guides for eyes
only.}

\end{figure*}
For all alloys in the studied composition range, the twinning stress
of Type 1 twins increases rapidly with decreasing temperature following
an universal dependence with the slope of about 0.04~MPa/K \cite{Heczko1.7K,StrakaT1xT2}.
This dependence is displayed in Fig.~\ref{fig:comparison_TS}a by
open blue diamonds (alloy 5) and small filled red squares (various
alloys from \cite{StrakaT1xT2} with $x=2.7-3.9$), and is labeled
as {}``Type 1 twins''. The microstructural model by Seiner et al.
\cite{Seiner_Microstructural} suggests that the increase originates
from the $a/b$-lamination (\{110\} compound twins) and thus it is
related to the difference between the $a$ and $b$ lattice constants
$(a-b)$. Alternatively, it can originate from modulation domains
(\{100\} compound twins) and thus it is related to angle $\gamma$,
or, more precisely, to $\gamma-90^{\circ}$. 

The determined $(a-b)$ as a function of relative temperature ($T-A_{S}$)
is given in Fig.~\ref{fig:comparison_TS}b. In spite of some scatter,
it is obvious from the figure that the $(a-b)$ dependences differ
significantly for different alloys. The higher is the transformation
temperature (or Mn content of the alloy or electron per atom concentration
$e/a$), the larger is the $(a-b)$ difference and it grows more rapidly
with the decreasing temperature. For alloys with transformation close
to room temperature (alloys 4 and 5), the $(a-b)$ difference is nearly
zero or zero in most of the temperature intervals studied, see also
Fig.~\ref{fig:a_b_summary}e, f. 

The significantly different $(a-b)$ dependences in different alloys,
Fig.~\ref{fig:comparison_TS}b, compared with the same universal
dependence of twinning stress for Type 1 twins, Fig.~\ref{fig:comparison_TS}a,
indicate that the increase in twinning stress cannot originate from
the $a/b$ lamination. Especially for alloy 5, the $a$ and $b$ are
very close to each other or identical resulting in no $a/b$-laminate,
but the twinning stress increase is about the same as in other alloys
(note that incorrect $a,b$ constants were listed in Ref.~\cite{Heczko1.7K}
due to an unnoticed typo). Thus, this experiment excludes the $a/b$-lamination
as the primary origin of the twinning stress increase for Type 1 twins. 

In contrast, better correlation is obtained with the $\gamma(T-A_{S})$
dependences. All alloys exhibit similar $\gamma(T-A_{S})$ dependences
in the temperature interval between $A_{S}$ and at least $A_{S}-50$~K,
Fig.~\ref{fig:comparison_TS}c. That compares well, within the experimental
scatter, with the observed universal dependence of twinning stress
of Type 1 twins, Fig.~\ref{fig:comparison_TS}a. This suggests that\textbf{
}the increase in twinning stress may originate from the $\gamma-90^{\circ}$
distortion. According to the theoretical model \cite{Seiner_Microstructural}
and experimental investigations \cite{Chulist_ab_laminate}, the propagating
Type 1 twin boundary interacts strongly with modulation domains (\{100\}
compound twins). The modulation domains may be distributed in bulk
or may be formed in the vicinity of the propagating boundary \cite{Chulist_ab_laminate}.
Larger $\gamma-90^{\circ}$ means that more energy is needed to form,
overcome or redistribute the modulation twins, so the positive correlation
between $\gamma$ and twinning stress is expected \cite{Seiner_Microstructural}.

Moreover, the $c$ lattice parameter or $c/a$ ratio exhibits similar
dependence in all alloys, Figs.~\ref{fig:c_constant} and \ref{fig:comparison_TS}d.
This can be significant because the $c/a$ ratio represents the twinning
shear, which must somehow influence the twinning stress. For example
in doped NM martensite the twinning stress decreased about tenfold
when c/a was reduced by about 5\% \cite{Sozinov_NM}. Thus, the observed
increase of twinning stress with decreasing temperature may be potentially
linked to the changes in $c$ or $c/a$. Nonetheless, $c/a$ as a
function of $(T-A_{S})$ is slightly different in different alloys,
Fig.~\ref{fig:comparison_TS}d, and its correlation with twinning
stress is slightly less convincing than for the case of $\gamma$.

\subsection{\label{sub:3.7.Relation-betwen-lattice}Relation between lattice
parameters and twinning stress for Type 2 twins}

The temperature dependences of twinning stress for Type 2 twins are
given in Fig.~\ref{fig:comparison_TS}a for each alloy separately
and additionally the observations for various alloys from \cite{StrakaT1xT2}
with $x=2.7-3.9$ are given as small filled green squares. The dependences
are labeled as {}``Type 2 twins'' in the figure. The twinning stress
is about constant between $A_{S}$ and some (low) temperature, below
which it rises rapidly. This temperature depends on alloy composition
and was found to coincide with the equilibrium temperature $T_{0}=(T_{IMT}+T_{RIMT})/2$,
which suggests that the twinning stress rise is related to the emerging
embryos of the 14M phase \cite{Straka_IMT}. Alternatively it was
suggested that the rise may also originate from changes in the lattice
constants and thus we compare here the lattice constants and twinning
stress. The comparison can be made only for alloys 1 and 2 and partly
for alloy 3; the rest of alloys exhibit the increase in twinning stress
below the measured temperature range.

No systematic correlation can be seen between the Type 2 twinning
stress increase and changes in lattice constants, Fig.~\ref{fig:comparison_TS}.
No significant changes in lattice parameters of alloy 1 occur at $T_{0}$
where the twinning stress starts rising. In contrast, alloy 2 exhibits
sudden changes in $a,b$ lattice parameters near $T_{0}$. Nonetheless,
alloy 3 shows similar sudden changes in lattice parameters far above
the $T_{0}$, with no impact on the twinning stress. Thus, there is
no clear correlation with the lattice constants, and the emerging
embryos of the 14M phase remain to be the most suspected reason for
increasing twinning stress of Type 2 twins.

\section{Conclusions}

The temperature dependences of lattice parameters $a,b,c$, and $\gamma$
were determined for Ni$_{50}$Mn$_{25+x}$Ga$_{25-x}$ single crystals
with 10M structure exhibiting very low twinning stress and magnetically
induced reorientation (MIR). With decreasing temperature, the lattice
parameters $a$ and monoclinic angle $\gamma$ increased, $c$ decreased,
while $b$ was nearly constant. Sudden large changes of lattice parameters
indicate the intermartensite transformation sequence 10M$\leftrightarrow$14M$\leftrightarrow$NM.
Additionally, in alloys with $x\leq3.5$, we observed small sudden
changes in $a,b$ lattice parameters (but not in $c$ parameter) far
above the intermartensite transformation temperature. This suggests
some fine structural rearrangement of 10M martensite, which may be
related to the refinement of twin structure on nanoscale. 

The direct comparison of the determined temperature dependences of
lattice parameters with the temperature dependence of twinning stress
indicate the following:
\begin{itemize}
\item Twinning stress of Type 1 twin boundaries is not correlated with $(a-b)$,
but it is reasonably correlated with $\gamma$, and there is also
a reasonable correlation with $c$ or $c/a$.
\item Twinning stress of Type 2 twin boundaries is not correlated with any
of the studied lattice parameters.
\end{itemize}
Thus, in contrast with the microstructural model \cite{Seiner_Microstructural},
the twinning stress of Type 1 twin boundaries does not depend significantly
on $a/b$ lamination. On the other hand, an alternative suggestion
of the model that $\gamma$ controls the twinning stress of Type 1
twin boundaries is in agreement with our experiment. The observed
correlation with $c/a$ may be also relevant \cite{Sozinov_NM} and
should be considered in the future models of twinning stress and MIR.

\section*{Acknowledgment}

This work was funded by the Academy of Finland, Czech Scientific Foundation
project of excellence No. 14-36566G, and by Academy of Sciences of
the Czech Republic in grant for international cooperation No. M100101241.
K.R. thanks for support of CSF grant No. P107/11/0391. The authors
thank Adaptamat Ltd. for providing samples for investigation and to
Alexei Sozinov for fruitful discussions.


\begin{thebibliography}{52}
\bibitem{Webster} Webster PJ, Ziebeck KRA, Town SL, Peak MS. Phil.
Mag. B 49 (1984) 295.

\bibitem{Ullakko1996} Ullakko K, Huang JK, Kantner C, O\textquoteright{}Handley
RC, Kokorin VV. Appl Phys Lett 69 (1996) 1966.

\bibitem{Buschov_book} Söderberg O, Sozinov A, Ge Y, Hannula S-P,
Lindroos VK. Giant magnetostrictive materials, in Buschow J (ed.).
Handbook of Magnetic Materials, Elsevier Science, Amsterdam 16 (2006)
1.

\bibitem{MSMPhenomena} Heczko O, Scheerbaum N, Gutfleisch O. Magnetic
shape memory phenomena in: Liu JP, Fullerton E, Gutfleisch O, Sellmyer
DJ (Eds.). Nanoscale Magnetic Materials and Applications, Springer
US (2009) 399.

\bibitem{Wilson_Review} Wilson SA, Jourdain RPJ, Zhang Q, Dorey RA,
et al. Mat Sci Eng R: Reports 56 (2007) 1.

\bibitem{likhachev2006} Likhachev AA, Sozinov A, Ullakko K. Mech
Mat 38 (2006) 551.

\bibitem{Sozinov_NM} Sozinov A, Lanska N, Soroka A, Zou W. App Phys
Lett 102 (2013) 021902.

\bibitem{Sozinov_14M} Sozinov A, Likhachev AA, Lanska N, Ullakko
K, App Phys Lett 80 (2002) 1746. 

\bibitem{karaca2006} Karaca HE, Karaman I, Basaran B, Chumlyakov
YI, Maier HJ. Acta Mat 54 (2006) 233.

\bibitem{Straka_2011Jphys} Straka L, Hänninen H, Soroka A, Sozinov
A. J Phys: Conf Ser 303 (2011) 012079. 

\bibitem{Schmidt} Schmidt H, J Phys: Conf Ser 303 (2011) 012078.

\bibitem{Benedict_2012} Schlüter K, Holz B, Raatz A, Adv Eng Mat
14 (2012) 682.

\bibitem{sehitoglu} Wang J, Sehitoglu H. Acta Mat 61 (2013) 6790.

\bibitem{Straka_IMT} Straka L, Sozinov A, Drahokoupil J, Kopecký
V, Hänninen H, Heczko O. J App Phys 114 (2013) 063504.

\bibitem{Kellis001MPa} Kellis D, Smith A, Ullakko K, Müllner P. J
Crystal Growth 359 (2012) 64.

\bibitem{Jaswon_Type_2} M. A. Jaswon and D. B. Dove, Acta Crystallogr.
13 (1960) 232.

\bibitem{Bilby_Type_2} B. A. Bilby, A. G. Crocker, Proc. R. Soc.
Lond. A 288 (1965) 240.

\bibitem{Mogylny} Mogylnyy G, Glavatskyy I, Glavatska N, Soderberg
O, Ge Y, Lindroos VK. Scripta Mat 48 (2003) 1427.

\bibitem{Nishida_2008} Nishida M, Hara T, Matsuda M, Ii S, Mat Sci
Eng A 481\textendash{}482 (2008) 18.

\bibitem{Sozinov_Type2} Sozinov A, Lanska N, Soroka A, Straka L,
App Phys Lett 99 (2011) 124103.

\bibitem{StrakaT1xT2} Straka L, Soroka A, Seiner H, Hänninen H, Sozinov
A. Scripta Mat 67 (2012) 25.

\bibitem{Heczko1.7K} Heczko O, Kopecký V, Sozinov A, Straka L, App
Phys Lett 103 (2013) 072405.

\bibitem{faran2011} Faran E, Shilo D. J Mech Phys Sol 59 (2011) 975.

\bibitem{Kaufman_Modulated_Mart} Kaufmann S, Niemann R, Thersleff
T, Rößler UK, Heczko O, Buschbeck J, Holzapfel B, Schultz L, Fähler
S. New J Phys 13 (2011) 053029.

\bibitem{Heczko_Acta2013} Heczko O, Straka L, Seiner H. Acta Mat
61 (2013) 622.

\bibitem{Salje1} Salje EKH. Phase Transitions 83 (2010) 657. 

\bibitem{Salje2} Lee WT, Salje EKH, Goncalves-Ferreira L, Daraktchiev
M, Bismayer U. Phys Rev B 73 (2006) 214110.

\bibitem{rajasekhara} Rajasekhara S, Ferreira PJ. Scripta Mat 53
(2005) 817.

\bibitem{faran2013} Faran E, Shilo D. J Mech Phys Sol 61 (2013) 726.

\bibitem{Lanska_2004} Lanska N, Soderberg O, Sozinov A, Ge Y, Ullakko
K, Lindroos VK. J App Phys 95 (2004) 8074.

\bibitem{Pagounis_abc}Pagounis E, Chulist, Szczerba MJ, Laufenberg
M, Appl. Phys. Lett. 105 (2014) 052405.

\bibitem{Glavatska_icomat} Glavatska N, Mogilniy G, Glavatsky I,
Danilkin S, Hohlwein D, Beskrovnij A, Söderberg O, Lindroos VK. J.
Phys IV France 113 (2003) 963.

\bibitem{Glavatsky_magnetic} Glavatskyy I, Glavatska N, Urubkov I,
Hoffman J-U, Bourdarot F, Mat Sci Eng A 481\textendash{}482 (2008)
298.

\bibitem{Seiner_Microstructural} Seiner H, Straka L, Heczko O, J
Mech Phys Sol 64 (2014) 198.

\bibitem{ShiloMatSciTech14}Faran E, Shilo D, Mat Sci Tech 30 (2014)
1545.

\bibitem{Straka_Acta_2011} Straka L, Heczko O, Seiner H, Lanska N,
Drahokoupil J, Soroka A, Fähler S, Hänninen H, Sozinov A, Acta Mat
59 (2011) 7450.

\bibitem{Chulist_ab_laminate} Chulist R, Straka L, Lanska N, Soroka
A, Sozinov A, Skrotzki W, Acta Mat 61 (2013) 1913.

\bibitem{Prevey} Prevéy PS. Adv X-Ray Anal 29 (1986) 103.

\bibitem{Book_XRD} Cullity BD, Stock SR. Elements of X-ray Diffraction,
third ed., Upper Saddle River, NJ:Prentice Hall (2001).

\bibitem{Heczko_MRB} Heczko O, Kope\v{c}ek J, Straka L, Seiner H,
Mat Res Bull 48 (2013) 5105.

\bibitem{Chulist_modulation} Chulist R, Straka L, Sozinov A, Lippmann
T, Skrotzki W. Scripta Mat 68 (2013) 671.

\bibitem{segui2005} Segu\'{i} C, Chernenko VA, Pons J, Cesari E,
Khovailo V, Takagi T. Acta Mat 53 (2005) 111.

\bibitem{segui2007} Segu\'{i} C, Pons J, Cesari E, Acta Mat 55 (2007)
1649.

\bibitem{Chernenko_Stress_temp} Chernenko VA, Pons J, Cesari E, Ishikawa
K, Acta Mat 53 (2005) 5071.

\bibitem{Cakr_IMT_2013} Çakr A, Righi L, Albertini F, Acet M, Farle
M, Aktürk S, J Appl Phys 114 (2013) 183912.

\bibitem{Kaufman_Adaptive_PRL} Kaufmann S, R\"{o}ßler UK, Heczko
O, Wuttig M, Buschbeck J, Schultz L, F\"{a}hler S, Phys Rev Lett
104 (2010) 145702.

\bibitem{Kim_X_Phase} Kim J-H, Fukuda T, Kakeshita T, Scripta Mat
54 (2006) 585.

\bibitem{Kushida_X_Phase} Kushida H, Hata K, Fukuda T, Terai T, Kakeshita
T, Scripta Mat 60 (2009) 96.

\bibitem{Ge_coarsening} Ge Y, Zarubova N, Heczko O, Hannula S-P,
{}``TEM in-situ observation of stress induced transition from modulated
to non-modulated martensite in Ni-Mn-Ga alloy'', submitted to Acta
Materialia, 2014.

\bibitem{Ustinov} Ustinov A, Olikhovska L, Glavatska N, Glavatskyy
I, J Phys: Conf Ser 226 (2010) 012016.

\bibitem{Ustinov_2} Olikhovska L, Ustinov A, Glavatska N, Glavatskyy
I, Esomat 2009 (2009) 02025. DOI:10.1051/esomat/200902025.

\bibitem{Righi_MSF} Righi L, Albertini F, Paoluzi A, Fabbrici S,
Villa E, Calestani G, Besseghini S. Mat Sci Forum 635 (2010) 33.
\end{thebibliography}
\end{document}